%acmacro
%boldgreek
%boxmacro
%fontmacro
%footmacro
%hyphen.tex
%mathmacro
%newmacro-l-n
%nonummacro
%sectmacro.bf
%theormacro.bf
%vecmacro.rm

\font\bfmit=cmmib10

\def\bfeta{\hbox{\bfmit\char'21}}

\def\bxi{\hbox{\bfmit\char'30}}

\def\bTheta{\hbox{\bf\char'2}}

\def\bSigma{\hbox{\bf\char'6}}

\global\def\squarebox{\vrule width0.5pt height5.6pt depth0pt\vbox{\hrule width5pt
height0.5pt depth0pt \vskip4.55pt \hrule width5pt height0.5pt depth0pt}\vrule width0.5pt
height5.6pt depth0pt}

\def\afoot{\strut\egroup}

\def\footnote#1{\let\sf=\empty
\ifhmode\edef\sf{\spacefactor=\the\spacefactor}\/\fi
#1\sf
\insert\footins\bgroup\eightpoint
\interlinepenalty100 \let\par=\endgraf
\leftskip=0pt \rightskip=0pt
\splittopskip=10pt plus 1pt minus 1pt \floatingpenalty=20000
\smallskip\item{#1}\bgroup\strut\aftergroup\afoot\let\next}
\skip\footins=12pt plus 2pt minus 4pt
\dimen\footins=50pc
\hyphenation{
Ac-ca-de-mia
ac-ce-le-ra-zio-ne
a-do-pe-ra-te
al-me-no
al-te-ra-re
al-tra
a-mi-co
am-met-te
an-no-ve-rar-si
an-ti-sim-me-tri-co
an-zi-tut-to
ap-pli-ca-zio-ne  ap-pli-ca-zio-ni
as-se-gnan-do as-se-gna-to
ba-se
ca-no-ni-co
ca-rat-te-ri-sti-co ca-rat-te-ri-sti-ci ca-rat-te-ri-sti-ca ca-rat-te-ri-sti-che
com-ples-so com-ples-si com-ples-sa com-ples-se
de-sti-na-to
di-sci-pli-na
e-la-bo-ra-zio-ne
ca-rat-te-riz-za-re ca-rat-te-riz-za-to ca-rat-te-riz-za-ta 
                    ca-rat-te-riz-za-ti ca-rat-te-riz-za-te
cer-ta-men-te
cia-scu-no  cia-scu-na
col-la-bo-ra-zio-ne
co-min-cia-mo
chi-mi-ca
com-por-ta
con-di-zio-ne
co-no-scen-za  co-no-scen-ze
con-si-de-ra-zio-ne con-si-de-ra-zio-ni
con-si-de-rar-si
con-si-de-ri
con-te-nu-ti
con-tro-va-rian-za
con-tro-va-rian-ti
Co-rio-lis
co-va-rian-ti
cre-scen-ti
de-di-ca-to de-di-ca-ti de-di-ca-ta de-di-ca-te
de-fi-ni-zio-ne
de-fi-ni-zio-ni
de-gli
del-la  del-le
det-ta-glia-to det-ta-glia-ti det-ta-glia-ta det-ta-glia-te 
di-mo-stra-zio-ne
di-sper-sio-ne
di-spo-si-zio-ni
di-stin-to di-stin-ta di-stin-ti di-stin-te
e-la-sti-ci-ty
e-la-sto-di-na-mi-ca  
e-li-mi-na-re
e-mi-sim-me-tri-co
e-pi-ste-mo-lo-gi-ci
e-qui-va-len-te e-qui-va-len-ti
e-qui-va-len-za
e-spe-rien-za e-spe-rien-ze
e-spli-ci-ta-to e-spli-ci-ta-ti e-spli-ci-ta-ta e-spli-ci-ta-te
e-spri-me-re
Es-sen-do
e-ster-no e-ster-ni e-ster-na e-ster-ne
e-ste-ti-ca 
eu-cli-deo
e-vo-lu-zio-ne
fe-no-me-ni
flui-do-di-na-mi-co
for-ma-li-smo
fra-zio-na-men-to
fun-zio-ne
ge-ne-ra-le   ge-ne-ra-li
ge-ne-ri-co ge-ne-ri-ca ge-ne-ri-ci ge-ne-ri-che
ge-nui-na-men-te
glo-ba-le
glo-bal-men-te
gno-seo-lo-gi-co gno-seo-lo-gi-ca
il-lu-stra-re il-lu-stra-to il-lu-stra-ta il-lu-stra-ti il-lu-stra-te
im-me-dia-ta-men-te
i-nol-tre
in-se-gna-re
in-ter-val-lo
in-tro-dot-to in-tro-dot-ti in-tro-dot-ta in-tro-dot-te
in-ver-sio-ne in-ver-sio-ni
in-ver-ti-bi-le
iper-su-per-fi-cie
iso-mor-fi-smo
iso-tro-pa i-so-tro-po i-so-tro-pi i-so-tro-pe  iso-tro-pia
i-stan-te i-stan-ti
li-nea-riz-za-zio-ne
lin-gui-sti-co lin-gui-sti-ca lin-gui-sti-ci lin-gui-sti-che
lo-gi-co lo-gi-ca lo-gi-ci lo-gi-che
ma-gne-to-e-las-to-dy-na-mics
ma-neg-ge-vo-le
ma-te-ma-ti-ca ma-te-ma-ti-ca-men-te ma-te-ma-ti-co
ma-te-ma-ti-che ma-te-ma-ti-ci
me-to-do-lo-gi-co me-to-do-lo-gi-ci
me-to-do-lo-gi-ca me-to-do-lo-gi-che
mi-nu-to mi-nu-ti 
ne-ces-sa-ria
ne-ces-sa-ria-men-te
new-to-nia-na
no-no-stan-te
no-stro no-stra no-stri no-stre
nul-la
on-de
o-pi-nio-ne o-pi-nio-ni
o-rien-ta-re
or-to-go-na-le
or-to-nor-ma-le
os-ser-via-mo
ov-via-men-te
pa-ra-me-tri-che
per-ce-zio-ne  per-ce-zio-ni
pleo-na-sti-co pleo-na-sti-ca
pos-sia-mo
po-sto
po-stu-la-to po-stu-la-ti po-stu-la-ta po-stu-la-te
pre-ce-den-te
pre-ci-sa-re
pro-ble-ma  pro-ble-mi
pro-ce-di-men-to
pro-po-si-to
pun-tua-li  
qual-sia-si
que-sto
rac-chiu-der-si
ra-dia-le
raf-fron-to
ra-gio-na-men-to
rap-pre-sen-ta-re
Ri-cor-dan-do
ri-guar-dar-si
ri-guar-di-no
ri-sco-per-to
sce-glie-re-mo
se-co-lo se-co-li
se-con-do
se-guen-ti
se-pa-ra-zio-ne
si-mul-ta-nea-men-te
si-ste-ma si-ste-mi
si-tua-zio-ne si-tua-zio-ni
sod-di-sfat-ta
sod-di-sfat-te
so-lu-zio-ne
so-sti-tui-sce 
sta-zio-na-rie
stes-sa
stes-se
stes-si
stes-so
stu-diar-ne
tem-pe-ra-tu-ra
tem-pu-sco-lo
tra-sfe-ri-men-to
tra-sfor-ma-zio-ni
tra-sver-sa 
u-gua-glian-za
u-ti-liz-zato
va-lo-re va-lo-ri
ve-ra-men-te
ve-ri-fi-ca
 }

\global\def\reale{\hbox{{\rm I}$\!${\rm R}}}

\global\def\lim{\mathop{\hbox{\rm lim}}}

\global\def\invh#1{#1^{\!\!\!\!{\raise4.1pt\hbox{$\scriptscriptstyle{-1}$}}}} 

\global\def\inv#1{#1^{\!\!\!\!{\raise3.1pt\hbox{$\scriptscriptstyle{-1}$}}}} 

\global\def\intr#1{#1^{\!\!\!{\raise4pt\hbox{$\scriptscriptstyle\circ$}}}}

\global\def\intrh#1{#1^{\!\!\!{\raise7pt\hbox{$\scriptscriptstyle\circ$}}}}

\magnification=1000
\baselineskip=18pt
\hsize=14truecm
\vsize=23truecm
\hoffset=1.5em
\voffset=18pt

\nopagenumbers
\headline={\centerline{\bf\folio}}
\outer\def\beginsection#1\par{\vskip 0pt plus.18\vsize\penalty-250
\vskip 0pt plus.18\vsize\medskip\vskip\parskip
\message{#1}\vbox{\noindent\hangindent=1.3pc\hangafter=1\bf#1}\nobreak\bigskip\noindent}
\outer\def\proclaim #1. #2\par{\medbreak
\noindent{\bf#1. \enspace}{\sl#2}\par
\ifdim\lastskip<\medskipamount \removelastskip\penalty55\medskip\fi}
\global\def\vt#1{\hbox{\bf#1}}

%\global\def\svt#1{\hbox{\sbf#1}}

%\global\def\ep{\pi(\nabla\hbox{\bf u})}

%\global\def\et{\eta(\hbox{\bf u})}

%\global\def\tr{\hbox{\bf s}(\hbox{\bf u})\hbox{\bf n}}

\centerline{THEORIES OF SPACE AND TIME}
\centerline{ COMPATIBLE WITH THE INERTIA PRINCIPLE}
\hfill\par\hfill\par
\centerline{Ermenegildo Caccese}
\centerline{Vito Antonio Cimmelli}
\centerline{Angelo Raffaele Pace}\par\hfill\par
\centerline{\it Department of Mathematics}
\centerline{\it University of Basilicata}
\centerline{\it 85100 POTENZA -- ITALY}\par\hfill\par\hfill\par
\vbox{\baselineskip=12pt\noindent 
{\bf Abstract}. {\rm A general formal definition of a theory of space and time compatible
with the inertia principle is given. The formal definition of reference frame and
inertial equivalence between reference frames are used to construct the class of
inertial frames. Then, suitable cocycle relations among the coefficients of
space-time transformations between inertial frames are established. The kinematical
meaning of coefficients and their reciprocity properties are discussed in some detail.
Finally, a rest frame map family is introduced as the most general constitutive
assumption to obtain the coefficients and to define a theory of space and time. Four
meaningful examples are then presented.}}\vskip1cm\par\noindent
{\bf 1. Introduction.}\vskip 0.5cm\par
The inertia principle (i.p.) is the basic postulate of all those descriptions of the
physical world in which the large scale gravitational forces are neglected and the uniform
motion of free particles have an ``absolute" meaning.\par
In its traditional formulation, the i.p. asserts that an {\it ``absolute" frame of
reference} exists with respect to which free particles have a constant speed [1], [13], [17].
However, the i.p. alone is not sufficient to determine the absolute frame of reference. From
the theoretical point of view, that last may be determined owing to some additional
assumptions which, unfortunately, fail when compared with the experiments. For instance,
Newton postulated the existence of absolute space and time on the ground of its
{\it rationalistic theology}. Descartes, and later Leibniz, supposed that space and time are
not {\it real things} and that space and matter coincide [1], [6-8]. On the other hand, in many
theories of electricity and magnetism of XIX century, the absolute space were identified
with the system respect to which the {\it aether} is at rest [17]. In that system, light
propagates isotropically. Finally, in the second half of XX century, many authors identified as
absolute the frame of reference respect to which the cosmic background radiation is
isotropic [4], [9], [22], [28].\par
Anyway, the i.p., alone, determines a whole class of reference frames, called the {\it
inertial frames of reference} (i.f.r.), which move, one respect to each other, with a constant
dragging velocity. Therefore, the problem of determining the mathematical structure of {\it
space-time transformations} between the frames of that class arises. That problem has been
considered by the Neopositivism as well as by other different phylosophic tendencies [2], [3],
[10-12], [15], [16]. It is universally accepted that a rigorous foundation of a theory of
space and time requires:\par\noindent
\item{a.} some operative criteria for the definition of space and time measurements,
togheter with an operative criterion for the synchronization of distant clocks, in any
i.f.r.;\par\noindent
\item{b.} a definition of the deformation rates of the measures of space and time, together
with a definition of the simultaneity defect, associated to any space-time transformation.
\par\noindent
Following that line of foundational research, the present work proposes a definition of a {\it
theory of space and time compatible with the inertia principle}, which is sufficiently
general to encompass all the examples known in the literature. These examples include both
those theories based on the principle of relativity (the Galilei and the Einstein-Poincar\'e
special relativity), and those based on the existence of an absolute frame, as for example
the  ``Lorentz relativity" [23] or the absolute theory, kinematically equivalent to special
relativity [22], [25-27], [29], [30].\par
The paper is organized as follows. In Section 2 the formal definition of a {\it frame of
reference} is introduced. As it is universally accepted in Epistemology [5], [10], [11], [19], in
order to coordinate the events by means of space and time, three independent operative
criteria must be available in any i.f.r.. The first two regulate the space and time
measurements, whereas the third regulates the synchronization of clocks. As far as that last
criterion is concerned, a distinction between synchronization by ``clock transport" and by
``first signals" has been made [5], [10], [13], [22], [25-27], [29], [30]. In both cases, a
definition which is operatively correct can be obtained assigning, by a convention, the
value of the ``one-way" velocities of the clocks carried as well as of the first signals.
Because of the motivations above, our formal definition of a frame of reference will assign,
together with a time $\Theta$ and a space $\Sigma$ (which are euclidean spaces of
dimension one and three), the set $C$ of the one-way velocities of material particles, which
is an open neighbourhood of the zero vector in the space of the translations of $\Sigma$:
$C\subseteq\bSigma$. The set $C$ is also important since it represents the domain of
definition of the constitutive relations which determine the coefficients of space-time
transformations as functions of the dragging velocity. Our definition is completed by the
assigning of the coordination map of the events: $\phi:M\to\Theta\times\Sigma$.\par
Section 3 is devoted to introduce the concept of {\it inertial equivalence} between frames of
reference. Two different frames of reference, $a$ and $b$, are said to be in the inertial
equivalence if the space-time transformation between them,
$\phi_{ba}=\phi_b\circ\phi_a^{-1}$, preserves the uniform motions. Inertial equivalence
between $a$ and $b$ implies that the the space-time transformation $\phi_{ba}$ is an
isomorphism between the affine structures of the relative times and spaces. Hence, this
transformation is determined once the dragging velocity (for instance, of $b$ with respect
to $a$) together with three additional coefficients are given: a positive real number
$\Delta_{ba}$; a linear form defined on the space relative to $a$,
$\tau_{ba}\in\bSigma_a^*$; a linear isomorphism between the relative spaces,
$\sigma_{ba}:\bSigma_a\to\bSigma_b$. We will show that the inertial equivalence is a
relation of equivalence in the sense of set theory. Therefore, from the formal point of view,
to assume i.p. is equivalent to fix a class of inertially equivalent frames of reference, $I$.
Finally, we prove that the coefficients of the space-time transformations between frames of
a same class, satisfy suitable {\it cocycle relations}.\par
In Section 4 we discuss the kinematical meaning of the coefficients of any space-time
transformation, $\phi_{ba}$, between inertially equivalent frames. The coefficient
$\Delta_{ba}$ determines the {\it deformation ratio of time measures}, performed with a
stationary clock in $a$, and with the clocks which are on its trajectory in $b$. The
coefficient $\tau_{ba}$ determines the {\it simultaneity defect}: two events which are
simultaneous in $a$, and spatially separated by a vector ${\vt r}\in\bSigma_a$, are no
longer simultaneous in $b$, but separated in time by the interval:
$$t^\prime_2-t^\prime_1=<\tau_{ba}\mid{\vt r}>.$$ The third coefficient, $\sigma_{ba}:
\bSigma_a\to\bSigma_b$, represents the spatial isomorphism which associates to any
stationary rod in $b$, ${\vt r}^\prime\in\bSigma_b$, its instantaneous configuration in $a$,
${\vt r}\in\bSigma_a$, through the relation: $${\vt r}^\prime=\sigma_{ba}{\vt r}.$$
Therefore $\sigma_{ba}$ determines the {\it deformation ratio of the space measures}:
$$\lambda_{ba}({\vt r})=:{\mid\sigma_{ba}{\vt r}\mid\over{\mid{\vt r}\mid}}.$$ The
eigenvalues associated to $\sigma_{ba}$ are said the {\it principal deformation ratios} of
space measures, while the corresponding eigenvectors are said the {\it principal vectors}. By
a {\it principal basis} we mean any oriented, orthonormal basis consisting of principal
vectors. If we refer the space $\Sigma_a$ to a principal basis, then we get a matrix
representation of the space-time transformation $\phi_{ba}$ which is diagonal with respect
to the space coordinates.\par
Inertial equivalence ``per se" does not imply any relation between the principal bases, the
dragging velocity and the simultaneity defect associated to a space-time transformation.
However three properties of the coefficients, which are not a logical consequence of
inertial equivalence, are assumed to hold in the relativistic theories as well as in the
theories based on the absolute frame. The first class of theories use them to describe all the
space-time transformations, while the second to describe only the transformations between
the absolute system and any other inertial frame. These properties, which we call the {\it
reciprocity conditions}, and analyze in Section 5, can be expressed as follows:\par\noindent
\item{1)} the dragging velocity is a principal vector;\par\noindent
\item{2)} the deformation ratio of space measures is constant on the plane normal to the
dragging velocity;\par\noindent
\item{3)} the simultaneity defect vanishes on the plane normal to the dragging velocity.
\par\noindent
A consequence of the reciprocity conditions is that the principal bases and the
simultaneity defect depend only on the dragging velocity, i.e. only on the ``relative motion" of
the two frames involved in the transformation. Therefore, the transformation is determined
by the dragging velocity, ${\vt u}$, up to four real parameters: the deformation ratio of time
measures, $\Delta_{ba}$; the two principal deformation ratios of space measures,
associated to the dragging velocity and its orthogonal plane: $$\lambda=:\lambda_{ba}({\vt
u})\quad;\quad\mu=:\lambda_{ba}({\vt r})\quad,\quad{\vt r}\in({\vt u})^\perp;$$ and, finally,
the component of the simultaneity defect in the direction of ${\vt u}$:
$$\theta=:<\tau_{ba}\mid u^{-2}{\vt u}>.$$ So, only when the reciprocity conditions hold for a
space-time transformation $\phi_{ba}$, then we find the familiar coordinate expression:
$$\eqalign{t^\prime&=\Delta_{ba}t+u\theta x^1\cr {x^\prime}^1&=\lambda(x^1-ut)\cr
{x^\prime}^\alpha&=\mu x^\alpha\quad,\quad\alpha=2,3.\cr}$$ Another consequence of the
reciprocity conditions is their symmetry, that is, they hold for the transformation
$\phi_{ba}$ if and only if they hold for the inverse transformation,
$\phi_{ab}=\phi_{ba}^{-1}$ also. Although in literature on special relativity and its ``test
theories", conditions 1-3 are interpreted as a consequence of homogeneity and isotropy of
the physical space [13], [15], [18], [19], [22], [25-28], it is in our opinion that they
express, instead, the property that the space-time transformation could be determined as a
function of the relative motion only.\par
Once we defined the frame of reference, introduced the inertial equivalence and fixed a class
$I$, in order to construct a theory of space and time compatible with the inertia principle it
remains to determine the explicit functional form of the coefficients of the transformations
involving the frames inside the class $I$. To this end, the cocycle relations alone are not
sufficient since they express only the {\it compatibility conditions} for any explicit
assignment of the coefficients. In order to determine these last explicitly, we use the so
called {\it rest frame principle}. It states that for any fixed i.f.r. $a\in I$, any other frame in
the class $I$ is uniquely determined by its dragging velocity with respect to $a$. In other
words, all the frames of reference in $I$ may be described, in $a$, as the rest frames of free
material particles. The rest frame principle, even if not mentioned in this form, is used to
construct any type of space-time transformations we know. In fact, its obvious consequence
is that for a given i.f.r. $a\in I$, the coefficients of any space-time transformation
$\phi_{ba}$ are functions of the dragging velocity relative to $a$. In
Section 6, the coefficients of space-time transformations are obtained by assigning a {\it
rest frame map family}: $${\cal R}=:\{{\cal R}_a:C_a\to I/a\in I\}.$$ Any ${\cal R}_a\in{\cal
R}$ is defined on the set $C_a$ of the one-way velocities of free material particles relative
to $a$; moreover, it is a bijection such that, according to the rest frame principle,
for any $b\in I$ the vector: $${\vt u}=:{\cal R}_a^{-1}(b)$$ is the dragging velocity of $b$
with respect to $a$. Further, the elements of ${\cal R}$ verify suitable compatibility
conditions coming from the cocycle relations. This goal is obtained by requiring that for any
pair of i.f.r. $a,b\in I$, the corresponding rest frame mappings, ${\cal R}_a:C_a\to I$ and
${\cal R}_b:C_b\to I$, be related by the law of composition of velocities:
$${\cal R}_b^{-1}\circ{\cal R}_a({\vt v})={\sigma_{ba}({\vt v}-{\vt
u})\over{\Delta_{ba}-<\tau_{ba}\mid{\vt v}>}}.$$ The conditions of compatibility, expressed
in that form, may be used both to select the constitutive functions determining the
coefficients of {\it any} space-time transformation (like in the relativistic theories), and
to obtain such constitutive functions {\it once those relative to a given ``absolute" frame are
known} (like in the theories based on an absolute frame). Section 6 is closed by defining a
{\it theory of space and time compatible with the inertia principle} as a triplet: $${\cal
T}\equiv(M;I;{\cal R})$$ given by the universe of events, $M$; the class of i.f.r., $I$; and a
rest frame map family ${\cal R}$ based on $I$. We quote four meaningful examples in
order to show that our definition encompass all the known theories. These examples are:
Galileian relativity, Einstein-Poincar\'e's special relativity, Lorentz's relativity and the
absolute theory kinematically equivalent to special relativity.\par
Let us conclude with a few words on the mathematical form in which we
present our approach. Since the original work by Minkowski in 1908, special relativity has
been formulated by introducing a Lorentz metric directly on the universe of events
$M$, [13]. According to such a formulation, both the class $I$ of i.f.r. and the rest frame map
family ${\cal R}$, may be obtained by this Lorentz metric. Recently, this approach has been
extended to  Galileian relativity [14], [20], [21], [24]. This formulation is based on two
metrics defined on $M$, one for space and the other for time. At our knowledge it is not yet
clear if and how a similar intrinsic geometric point of view may be extended to the theories
with an absolute frame. In any case, as first noted by Poincar\'e [23], to these theories
cannot be associated a group of transformations. Because of the motivations above, we
formulated our approach by using a formalism which neither involves directly the universe
of events, $M$, nor tries to introduce an intrinsic geometric structure on it.\vskip
1cm\par\noindent
{\bf 2. Global space-time reference frames.}\vskip 0.5cm\par
Let $M$ be the set of events. We call $M$ the {\it universe}. A {\it global space-time
reference frame} of $M$ consists of a splitting of $M$ into time and space, together with the
specification of the set of one-way velocities of material particles, relative to an observer.
More formally, a reference frame of $M$ is a triplet, $$(\Theta\times\Sigma; C;\phi),$$
consisting of:\par\noindent
\item{(a)} a product, $\Theta\times\Sigma$, of an one-dimensional by a three-dimensional
affine spaces, endowed with orientations and euclidean metrics;\par\noindent
\item{(b)} an open convex neighbourhood of the zero vector, with regular boundary, in the
vector space of translations of $\Sigma$, $C\subseteq\bSigma$;\par\noindent
\item{(c)} a one-to-one map, $\phi:M\to\Theta\times\Sigma$.\par
$\Theta$ represents the {\it relative time}. Euclidean structure and orientation of $\Theta$
are fixed by a non zero vector in the vector space of translations, ${\vt
e}_0\in\bTheta$, which represents the oriented unit time interval. $\Sigma$ represents the
{\it relative space}. Euclidean metric of $\Sigma$ is fixed by an euclidean
scalar product in the vector space of translations, $h:\bSigma\times\bSigma\to\reale$. Any
normal vector, ${\vt e}\in\bSigma$, $h({\vt e},{\vt e})=1$, represents an oriented rod with
unitary lenght. The orientation of $\Sigma$ is fixed by a class of
$SO(3)$-related orthonormal bases of $\bSigma$.\par $C$ represents the set of {\it one-way
relative velocities} of material particles. It is assumed to be open in account of a general
``stability" of physical laws. This is tantamount to admit that physical laws are such that if
${\vt v}\in C$ is any physical velocity, then neighbourhoods of ${\vt v}$ can be found, all
consisting of physical velocities. Convexity of $C$ and smoothness of its boundary account
for a reasonable compatibility between the description of motion and the causality relation
in $M$.\par
Finally, $\phi$ represents the {\it global coordination map}, which associates
any event, $x\in M$, with an instant of time and a point of space: $\phi(x)=(T,P)$. From the
operational point of view, the definition of $\phi$ is subordinated to some fixed rules for
the measurements of time and space, together with a fixed rule for the synchronization of
clocks. Different rules give rise, in general, to different theories of space and time.
Space and time separations between events are defined in the obvious
way, by means of $\phi$; that is, if $\phi(x_1)=(T_1,P_1)$, and $\phi(x_2)=(T_2,P_2)$,
then:\par\noindent \item{-} the {\it space separation} between $x_1$ and $x_2$ is
represented by the vector ${\vt r}=:P_2-P_1\in\bSigma$, and its measure is given by the
real number: $\mid{\vt r}\mid=:h({\vt r},{\vt r})^{1/2}\geq 0$;\par\noindent
\item{-} the {\it time separation} between $x_1$ and $x_2$ is represented by the vector
$T_2-T_1\in\bTheta$, and its measure is given by the real number: $<{\vt e}_0^*\mid
T_2-T_1>$, where ${\vt e}_0^*\in\bTheta^*$ is the dual vector of ${\vt e}_0$.\par\noindent
The events $x_1$ and $ x_2$ are said to be {\it genidentic} [5], [10], if $P_1=P_2$; on the
other hand, they are said to be {\it simultaneous}, if $T_1=T_2$.\par
Once an origin of time, $T_0\in\Theta$, is fixed, then a {\it time scale} for the events
is defined within the reference frame. Indeed, for any $x\in M$, if $\phi(x)=(T,P)$, then:
$$T=T_0+t{\vt e}_0,$$ where $t=:<{\vt e}_0^*\mid T-T_0>$. Now, let $T_{0^\prime}=
T_0+t_{0^\prime}{\vt e}_0$ be a different time origin, then the two time scales
defined by $T_{0^\prime}$ and $T_0$ are related by: $t^\prime=t-t_{0^\prime}$. Moreover,
once an origin in space, $O\in\Sigma$, is fixed, then a {\it vector representation} of events
is defined within the reference frame. Indeed, if $\phi(x)=(T,P)$, then: $$P=O+{\vt r},$$ where
${\vt r}\in\bSigma$ is the vector which translates $O$ on $P$. Now, let $O^\prime=O+{\vt
r}_{O^\prime}$ be a different space origin, then the two vector representations defined by
$O^\prime$ and $O$ are related by: ${\vt r}^\prime={\vt r}-{\vt r}_{O^\prime}$. The singling
out of both $T_0$ and $O$, gives rise to a representation of events by means of their time
scale coordinates and space vectors, instead of the more formal corresponding objects in
$\Theta$ and $\Sigma$. In fact, for any $x\in M$, we have: $$\phi(x)=(T,P)=(T_0,O)+(t{\vt
e}_0,{\vt r}).$$\par
Let now ${\vt v}\in C$ be any velocity; then the {\it uniform motions} with velocity ${\vt v}$
are represented in $\Theta\times\Sigma$ by the straight lines: $$m(t)=:(T_0,O)+(t{\vt
e}_0,P_0-O+t{\vt v})=(T_0+t{\vt e}_0,P_0+t{\vt v}),$$ in which $T_0$ and $O$ are the
origins in time and in space, and $P_0$ is the starting point of the motion. More generally, a
motion is represented by a curve: $$m(t)=(T_0,O)+(t{\vt e}_0,{\vt r}(t)),$$ in which
$t\mapsto{\vt r}(t)$ is a $\bSigma$-valued function such that $\dot{\vt r}(t)\in C$. Velocity
and acceleration at the time $t$ are defined by: ${\vt v}(t)=:\dot{\vt r}(t)$ and ${\vt
a}(t)=:\ddot{\vt r}(t)$. Clearly, ${\vt v}(t)$ and ${\vt a}(t)$ are invariant with respect to
substitutions of the origin in time and space.\vskip 1 cm\par\noindent
{\bf 3. Inertial equivalence and space-time transformations.}\vskip 0.5cm\par
Let $a=:(\Theta_a\times\Sigma_a;C_a;\phi_a)$, and let $b=:(\Theta_b\times\Sigma_b;C_b;
\phi_b)$, be two reference frames. The {\it space-time transformation from $a$ to $b$} is
defined by: $$\phi_{ba}=:\phi_b\circ\phi_a^{-1}:\Theta_a\times\Sigma_a\to\Theta_b
\times\Sigma_b.$$ We say that $a$ and $b$ are {\it inertially equivalent}, and write
$a\sim_I b$, if the following properties are fulfilled by $\phi_{ba}$:\par\noindent
\item{(a)} $\phi_{ba}$ is a smooth diffeomorphism;\par\noindent
\item{(b)} there exist a  smooth function, $\delta_{ba}:C_a\to{\reale}^+$, and an orientation
preserving smooth diffeomorphism, $\Phi_{ba}:C_a\to C_b$, such that the following equation
holds: $$\phi_{ba}(T_0+t{\vt e}_0,P_0+t{\vt v})=(T^\prime_0+\delta_{ba}({\vt v})t
{\vt e}_0^\prime,P^\prime_0+\delta_{ba}({\vt v})t\Phi_{ba}({\vt v})),\eqno(3.1)$$
\item{} where $$(T_0^\prime,P_0^\prime)=:\phi_{ba}(T_0,P_0),$$
\item{} for any $(T_0,P_0,{\vt v})\in\Theta_a\times\Sigma_a\times C_a$, and for any
$t\in\reale$.\par\noindent
In other words, inertial equivalence means that $\phi_{ba}$ transforms uniform motions
relative to $a$ into uniform motions relative to $b$. Moreover, inertial equivalence requires
that if a material particle moves with constant velocity ${\vt v}\in C_a$ with respect to
$a$, then its velocity with respect to $b$ depends only on ${\vt v}$, by means of: ${\vt
v}^\prime=\Phi_{ba}({\vt v})$. Finally, if $x_1,x_2\in M$ is any pair of events connected by
the propagation of this particle, that is, if: $$\eqalign{\phi_a(x_i)&=(T_0+t_i{\vt
e}_0,P_0+t_i{\vt v})\cr \phi_b(x_i)&=(T^\prime_0+t^\prime_i{\vt e}^\prime_0,
P^\prime_0+t^\prime_i{\vt v}^\prime)\cr}\quad,\quad i=1,2,$$ then inertial equivalence
requires that the ratio of the time separation between $x_1$ and $x_2$ measured in $b$ by
that measured in $a$, depends only on ${\vt v}$ by means of:
$${t^\prime_2-t^\prime_1\over{t_2-t_1}}=\delta_{ba}({\vt v}).$$\par
Although the most natural way to define the inertial equivalence between the reference
frame $a$ and $b$ is to require that the space-time transformation $\phi_{ba}$ preserves
the uniform motions, it is a very easy consequence that $\phi_{ba}$ also preserves the affine
structures of time and space.\vskip 0.5cm\par\noindent
{\bf Theorem 1:} {\it If $a\sim_I b$, then the transformation $\phi_{ba}$ is an affine
isomorphism.}\par\noindent
Proof. Since $\phi_{ba}$ is a smooth diffeomorphism, then its differential is a smooth
function which takes its values in the set of linear isomorphisms between
$\bTheta_a\oplus\bSigma_a$ and $\bTheta_b\oplus\bSigma_b$:
$$d\phi_{ba}:\Theta_a\times\Sigma_a\to(\bTheta_a\oplus\bSigma_a)^*\otimes
(\bTheta_b\oplus\bSigma_b).$$ Now, if we represent $\phi_{ba}$ by the
matrix: $$d\phi_{ba}=\pmatrix{{\partial T^\prime \over{\partial T}}{\vt e}_0^*\otimes
{\vt e}_0^\prime&{\partial T^\prime\over{\partial P}}\otimes{\vt e}_0^\prime\cr
{\vt e}^*_0\otimes{\partial P^\prime\over{\partial T}}&{\partial
P^\prime\over{\partial P}}\cr}$$ then we get the functions:
$$\eqalign{{\partial T^\prime\over{\partial T}} &:\Theta_a\times\Sigma_a\to{\reale}\cr
{\partial T^\prime\over{\partial P}}&:\Theta_a\times\Sigma_a\to\bSigma^*_a\cr
{\partial P^\prime\over{\partial T}}&:\Theta_a\times\Sigma_a\to\bSigma_b\cr
{\partial P^\prime\over{\partial P}}&:\Theta_a\times
\Sigma_a\to\bSigma^*_a\otimes\bSigma_b,\cr}$$ which are smooth and, in addition,
$\partial P^\prime/\partial P$ takes its values in the set of orientation preserving linear
isomorphisms between $\bSigma_a$ and $\bSigma_b$. Now, by differentiating (3.1), we find:
$$\eqalign{\left({\partial T^\prime\over{\partial T}}\right)_{T_0,P_0}&+<\left({\partial
T^\prime\over{\partial P}}\right)_{T_0,P_0}\mid{\vt v}>=\delta_{ba}({\vt v})\cr
\left({\partial P^\prime\over{\partial T}}\right)_{T_0,P_0}&+\left({\partial
P^\prime\over{\partial P}}\right)_{T_0,P_0}{\vt v}=\delta_{ba}({\vt v})\Phi_{ba}({\vt
v}),\cr}\eqno(3.2)$$ for any $(T_0,P_0,{\vt v})\in\Theta_a\times\Sigma_a\times C_a$.
Now, if we set ${\vt v}={\vt 0}$ in (3.2), and define: $\Delta_{ba}=:\delta_{ba}({\vt 0})$;
${\vt u}_{ba}=:\Phi_{ba}({\vt 0})$, then we find that $\partial T^\prime/\partial T$ and
$\partial P^\prime/\partial T$ are constants: $${\partial T^\prime\over{\partial
T}}=\Delta_{ba}\quad;\quad{\partial P^\prime\over{\partial P}}=\Delta_{ba}{\vt u}_{ba}.$$
Finally, since the right-hand sides of (3.2) depend only on ${\vt v}\in C_a$, then $\partial
T^\prime/\partial P$ and $\partial P^\prime/\partial P$ also are constants. We denote their
values by: $$\tau_{ba}=:{\partial T^\prime\over{\partial P}}\quad;\quad
\sigma_{ba}=:{\partial P^\prime\over{\partial P}}.$$
We have just proved that $\phi_{ba}$ is an affine isomorphism, provided that its
differential is the constant linear function: $$d\phi_{ba}=\pmatrix{\Delta_{ba}{\vt
e}^*_0\otimes{\vt e}^\prime_0&\tau_{ba}\otimes{\vt e}^\prime_0\cr {\vt
e}^*_0\otimes\Delta_{ba}{\vt u}_{sr}&\sigma_{ba}\cr}.$$\hfill\squarebox\par\hfill\vskip
0.5cm\par
The space-time transformation connecting two inertially equivalent reference
frames is completely determined by the four quantities:\par\noindent
\item{1.} the number $\Delta_{ba}\in\reale^+$;\par\noindent
\item{2.} the covector $\tau_{ba}\in\bSigma_a^*$;\par\noindent
\item{3.} the vector ${\vt u}_{ba}\in\bSigma_b$;\par\noindent
\item{4.} the linear isomorphism $\sigma_{ba}:\bSigma_a\to\bSigma_b$;\par\noindent
we call them the {\it coefficients of the transformation} $\phi_{ba}$. Any other 
kinematical quantity associated with $\phi_{ba}$ can be expressed as a function of these
coefficients. For example, by (3.2) we derive: $$\eqalign{\delta_{ba}({\vt v})&=
\Delta_{ba}+<\tau_{ba}\mid{\vt v}>\cr \Phi_{ba}({\vt v})&=(\Delta_{ba}+<\tau_{ba}\mid{\vt
v}>)^{-1}(\Delta_{ba}{\vt u}_{ba}+\sigma_{ba}{\vt v}).\cr}\eqno(3.3)$$\par
The most important features of the inertial equivalence is that it divides the set of all the
reference frames of $M$ into equivalence classes and, moreover, in any of these classes
suitable {\it cocycle relations} hold among the coefficients of the space-time
transformations. Therefore, the inertia principle, mathematically, consists in singling
out an equivalence class of reference frames. Moreover, the cocycle relations can be
interpreted as the most general compatibility conditions for the determination of the
coefficients of space-time transformations in any theory of space and time . We now
determine the cocycle relations.\vskip 0.5cm\par\noindent
{\bf Theorem 2:} {\it The inertial equivalence is an equivalence relation in the set of all the
global space-time reference frames of $M$. Moreover, if $a,b,c$ are three frames belonging
to the same class, then the coefficients of $\phi_{ab},\phi_{ba},\phi_{cb}$, and $\phi_{ca}$,
fulfil the following cocycle relations:} $$\eqalign{(\Delta_{ab}+<\tau_{ab}\mid{\vt
u}_{ba}>)\Delta_{ba}&=1\cr \Delta_{ab}{\vt u}_{ab}+\sigma_{ab}{\vt u}_{ba}&=0\cr
\Delta_{ab}\tau_{ba}+\tau_{ab}\circ\sigma_{ba}&=0\cr
\sigma_{ab}\circ(\sigma_{ba}-\tau_{ba}\otimes{\vt u}_{ba})&=id_{\bSigma_a}\cr
(\Delta_{ba}+<\tau_{ba}\mid{\vt u}_{ab}>)\Delta_{ab}&=1\cr
\Delta_{ba}{\vt u}_{ba}+\sigma_{ba}{\vt u}_{ab}&=0\cr
\Delta_{ba}\tau_{ab}+\tau_{ba}\circ\sigma_{ab}&=0\cr
\sigma_{ba}\circ(\sigma_{ab}-\tau_{ab}\otimes{\vt u}_{ab})&=
id_{\bSigma_b}\cr}\eqno(3.4)$$
$$\eqalign{(\Delta_{cb}+<\tau_{cb}\mid{\vt u}_{ba}>)\Delta_{ba}&=\Delta_{ca}\cr
(\Delta_{cb}+<\tau_{cb}\mid{\vt u}_{ba}>)^{-1}(\Delta_{cb}{\vt u}_{cb}+\sigma_{cb}{\vt
u}_{ba})&=\Phi_{cb}({\vt u}_{ba})={\vt u}_{ca}\cr
\Delta_{cb}\tau_{ba}+\tau_{cb}\circ\sigma_{ba}&=\tau_{ca}\cr
\sigma_{cb}\circ(\sigma_{ba}-\tau_{ba}\otimes{\vt u}_{bc})&=\sigma_{ca}.\cr}\eqno(3.5)$$
Proof. The relation $\sim_I$ is manifestly reflexive, provided that $\phi_{aa}=id_{\Theta_a
\times\Sigma_a}$. Let's now assume $a\sim_I b$. Since $\phi_{ab}=\phi_{ba}^{-1}$, then
$\phi_{ab}$ is an affine isomorphism, and $d\phi_{ab}=(d\phi_{ba})^{-1}$; so if we set:
$$d\phi_{ab}=\pmatrix{\Delta_{ab}{\vt e}^{\prime*}_0\otimes{\vt e}_0&\tau_{ab}\otimes
{\vt e}_0\cr {\vt e}^{^\prime*}_0\otimes\Delta_{ab}{\vt u}_{ab}&\sigma_{ab}\cr},$$ then the
identities: $$d\phi_{ab}\circ d\phi_{ba}=id_{\bTheta_a\oplus\bSigma_a}\quad;\quad
d\phi_{ba}\circ d\phi_{ab}=id_{\bTheta_b\oplus\bSigma_b},$$ lead to the cocycle relations
(3.4). Moreover, for any uniform motion with respect to $b$, we find: $$\eqalign{\phi_{ab}
(T^\prime_0+t^\prime{\vt e}^\prime_0,P^\prime_0+t^\prime{\vt v}^\prime)&=
\phi_{ab}(T^\prime_0,P^\prime_0)+d\phi_{ab}(t^\prime{\vt e}^\prime_0,t^\prime{\vt
v}^\prime)=\cr &=(T_0+\delta_{ab}({\vt v}^\prime)t^\prime{\vt e}_0,
P_0+\delta_{ab}({\vt v}^\prime)t^\prime{\vt v}),\cr}$$ where:
$$\eqalign{(T_0,P_0)&=:\phi_{ab}(T^\prime_0,P^\prime_0),\cr
\delta_{ab}({\vt v}^\prime)&=:\Delta_{ab}+<\tau_{ab}\mid{\vt v}^\prime>,\cr
{\vt v}=:\Phi_{ab}({\vt v}^\prime)&=:(\Delta_{ab}+<\tau_{ab}\mid{\vt v}^\prime>)^{-1}
(\Delta_{ab}{\vt u}_{ab}+\sigma_{ab}{\vt v}^\prime).\cr}$$
Moreover, from (3.4) we derive that: $$\delta_{ab}({\vt
v}^\prime)={1\over{\delta_{ba}(\Phi_{ba}^{-1} ({\vt v}^\prime))}}\quad;\quad\Phi_{ab}({\vt
v}^\prime)=\Phi_{ba}^{-1}({\vt v}^\prime); \eqno(3.6)$$ hence, the function
$\delta_{ab}:C_b\to\reale$ is smooth and positive valued, and the function
$\Phi_{ab}:C_b\to\bSigma_a$ is an orientation preserving smooth diffeomorphism of $C_b$
onto $C_a$. This means that: $b\sim_I a$.\par\noindent Finally, let's assume that $a\sim_I
b$ and $b\sim_I c$. Since $\phi_{ca}=\phi_{cb}\circ\phi_{ba}$, then $\phi_{ca}$ is an affine
isomorphism; so if we set: $$d\phi_{ca}=\pmatrix{\Delta_{ca}{\vt e}^*_0\otimes{\vt
e}^{\prime\prime}_0& \tau_{ca}\otimes{\vt e}^{\prime\prime}_0\cr {\vt
e}^\prime_0\otimes\Delta_{ca}{\vt u}_{ca}&\sigma_{ca}\cr},$$ then the chain rule:
$d\phi_{ca}=d\phi_{cb}\circ d\phi_{ba}$, leads to the cocycle relations (3.5). Now, for any
uniform motion with respect to $a$, we find: $$\eqalign{\phi_{ca}(T_0+t{\vt e}_0,P_0+t{\vt
v})&=\phi_{ca}(T_0,P_0)+ d\phi_{ca}(t{\vt e}_0,t{\vt v})=\cr
&=(T^{\prime\prime}_0+\delta_{ca}({\vt v})t{\vt
e}^{\prime\prime}_0,P^{\prime\prime}_0+\delta_{ca}({\vt v})t{\vt v}^{\prime\prime}),\cr}$$
where: $$\eqalign{(T^{\prime\prime}_0,P^{\prime\prime}_0)&=:\phi_{ca}(T_0,P_0),\cr
\delta_{ca}({\vt v})&=:\Delta_{ca}+<\tau_{ca}\mid{\vt v})>,\cr {\vt
v}^{\prime\prime}=\Phi_{ca}({\vt v})&=:(\Delta_{ca}+<\tau_{ca}\mid{\vt
v})>)^{-1}(\Delta_{ca}{\vt u}_{ca}+\sigma_{ca}{\vt v}).\cr}$$ Moreover, from (3.5) we derive
that: $$\delta_{ca}({\vt v})=\delta_{cb}(\Phi_{ba}({\vt v}))\cdot\delta_{ba}({\vt v})
\quad;\quad\Phi_{ca}({\vt v})=\Phi_{cb}\circ\Phi_{ba}({\vt v});\eqno(3.7)$$ hence, the
function $\delta_{ca}:C_a\to\reale$ is smooth and positive definite, and the function
$\Phi_{ca}:C_a\to\bSigma_c$ is an orientation preserving $C^\infty$-diffeomorphism of
$C_a$ onto $C_c$. This means that $a\sim_I c$.\hfill\squarebox\par\hfill\vskip
1cm\par\noindent
{\bf 4. Kinematical meaning of the coefficients of a transformation.}\vskip 0.5cm\par
Let $a$ and $b$ be two space-time reference frames of $M$ which are inertially
equivalent. The coefficients of the space-time transformation $\phi_{ba}$ have the
kinematical meanings of dragging velocity, deformation ratios of time and space measures,
and simultaneity defect, affecting $\phi_{ba}$. To state this more clearly, we fix an
event, $o\in M$, in order to single out origins in time and space within $a$ and $b$, by:
$$\phi_a(o)=(T_0,O)\quad;\quad\phi_b(o)=(T^\prime_0,O^\prime).$$ Then, any event $x\in M$
is represented by its time and space coordinates by: $$\eqalign{\phi_a(x)&=(T,P)
=(T_0,O)+(t{\vt e}_0,{\vt r})\cr
\phi_b(x)&=(T^\prime,P^\prime)=(T^\prime_0,O^\prime)+(t^\prime{\vt e}^\prime_0,{\vt
r}^\prime),\cr}$$ and the space-time transformation, $(T^\prime,P^\prime)=
\phi_{ba}(T,P)$, is represented by the linear transformation:
$$\eqalign{t^\prime&=\Delta_{ba}t+<\tau_{ba}\mid{\vt r}>\cr
{\vt r}^\prime&=\Delta_{ba}t{\vt u}_{ba}+\sigma_{ba}{\vt r}.\cr}\eqno(4.1)$$\par
The vector ${\vt u}_{ba}$ represents the {\it dragging velocity} of the space $\Sigma_a$ with
respect to $b$. In other words, ${\vt u}_{ba}=\Phi_{ba}({\vt 0})\in C_b$ is the velocity,
measured in $b$, of any particle which is at rest in $a$. The dragging velocity ${\vt u}_{ba}$
is related to the dragging velocity ${\vt u}_{ab}$ of the space $\Sigma_b$ with respect to
$a$ by the cocycle relation $(3.4)_6$: $${\vt u}_{ba}=-\Delta_{ba}^{-1}\sigma_{ba}{\vt
u}_{ab}.$$ Therefore, the equations (4.1) may be also expressed as follows:
$$\eqalign{t^\prime&=\Delta_{ba}t+<\tau_{ba}\mid{\vt r}>\cr {\vt r}^\prime&=
\sigma_{ba}({\vt r}-t{\vt u}_{ab}),\cr}\eqno(4.2)$$ while the transformation rule for
velocities takes the form: $${\vt v}^\prime=\Phi_{ba}({\vt v})=(\Delta_{ba}+
<\tau_{ba}\mid{\vt v}>)^{-1}\sigma_{ba}({\vt v}-{\vt u}_{ab}).\eqno(4.3)$$ Finally, the
transformation rules for velocity and acceleration of arbitrary motions can be expressed by
means of the coefficients of $\phi_{ba}$. Indeed, let $m(t)=(T_0, O)+(t{\vt e}_0,{\vt r}(t))$
be any motion referred to $a$. The same motion, referred to $b$,
$m^\prime(t^\prime)=(T^\prime_0,O^\prime)+(t^\prime{\vt e}^\prime_0,{\vt
r}^\prime(t^\prime))$, is expressed as a function of the time scale of $a$ by means of:
$$\phi_{ba}\circ m(t)=m^\prime(f(t)),$$ where: $$f(t)=\Delta_{ba}t+<\tau_{ba}\mid {\vt
r}(t)>.$$ Now, performing the time derivatives, we find the transformation rules:
$$\eqalign{{\vt v}^\prime(t^\prime)\mid_{t^\prime=f(t)}&=(\Delta_{ba}+<\tau_{ba}\mid{\vt
v}(t)>)^{-1} \sigma_{ba}({\vt v}(t)-{\vt u}_{ab})=\Phi_{ba}({\vt v}(t))\cr {\vt
a}^\prime(t^\prime)\mid_{t^\prime=f(t)}&=(\Delta_{ba}+<\tau_{ba}\mid{\vt v}(t)>)^{-2}
(\sigma_{ba}{\vt a}(t)-<\tau_{ba}\mid{\vt a}(t)>\Phi_{ba}({\vt v}(t))).\cr}\eqno(4.4)$$
Accordingly to the inertia principle, from (4.4) it follows that the uniform motion of a
particle has an ``absolute meaning". Nevertheless, the acceleration vector itself is generally
frame-dependent.\par
Let now $x_1,x_2\in M$ be two events which are genidentic with respect to $a$ (cfr. sect. 2).
Then $x_1$ and $x_2$ are connected by the propagation of a particle which is at rest in $a$
at the point ${\vt r}_2={\vt r}_1$, hence:
$${t^\prime_2-t^\prime_1\over{t_2-t_1}}=\delta_{ba}({\vt o})=\Delta_{ba},$$ where
$t^\prime_2-t^\prime_1$ and $t_2-t_1$ measure the time separations between $x_1$ and
$x_2$ respectively in $b$ and in $a$. Hence, the coefficient $\Delta_{ba}\in\reale^+$
represents the {\it deformation ratio of time measures}, between $a$-genidentic events,
affecting the transformation $\phi_{ba}$.\par
Let's now assume that $x_1$ and $x_2$ are simultaneous with respect to $a$. Then
$t_2=t_1$ and, by (4.2), it follows that: $$t^\prime_2-t^\prime_1=<\tau_{ba}\mid{\vt
r}_2-{\vt r}_1>.$$ Hence, the  number $<\tau_{ba}\mid{\vt r}_2-{\vt r}_1>$
measures the time separation, with respect to $b$, between $x_1$ and $x_2$. Therefore, the
coefficient $\tau_{ba}\in\bSigma^*_a$ represents the {\it simultaneity defect} affecting
the transformation $\phi_{ba}$. The simultaneity of $x_1$ and $x_2$ is conserved by
$\phi_{ba}$ if and only if: ${\vt r}_2-{\vt r}_1\in$ ker $\tau_{ba}$; therefore we call ker
$\tau_{ba}\subseteq\bSigma_a$ the {\it subspace of conserved simultaneity} associated to
$\phi_{ba}$.\par
Now, we consider a rod which is at rest in $b$, and denote by ${\vt r}^\prime\in\bSigma_b$
the vector connecting its extreme points. Since these points are uniformly moving in $a$
with dragging velocity ${\vt u}_{ab}$, then we derive from (4.2) that their positions in
$a$, at any time, are connected by the vector ${\vt r}\in\bSigma_a$ determined by: ${\vt
r}^\prime=\sigma_{ba}{\vt r}$. Hence, for any rod which is at rest in $b$, $\sigma_{ba}$
maps its instantaneous configurations in $a$ onto its configuration in $b$. We call
$\sigma_{ba}$ the {\it space transformation} associated to $\phi_{ba}$.
From $\sigma_{ba}$ we derive the function: $$\lambda_{ba}({\vt r})=:{\mid{\vt
r}^\prime\mid\over{\mid{\vt r}\mid}}={h_b(\sigma_{ba}{\vt r},\sigma_{ba}{\vt
r})^{1/2}\over{h_a({\vt r},{\vt r})^{1/2}}},\eqno(4.5)$$ which represents the {\it deformation
ratio of space measures} affecting the transformation $\phi_{ba}$. Moreover, we define the
{\it principal deformation ratios of} $\phi_{ba}$ as the square roots of the eigenvalues of
the symmetric bilinear form $h_b\circ(\sigma_{ba}\times\sigma_{ba})$, with respect to
$h_a$. Finally, we call {\it principal vector of} $\phi_{ba}$, any eigenvector of
$h_b\circ(\sigma_{ba}\times\sigma_{ba})$. Clearly, a non zero vector, ${\vt
r}\in\bSigma_a$, is principal if and only if: $$\forall\bxi\in\bSigma_a\quad,\quad
h_b(\sigma_{ba}\bxi,\sigma_{ba}{\vt r})=\lambda_{ba}^2({\vt r})h_a(\bxi,{\vt r}).$$ We call
{\it principal basis} of the transformation $\phi_{ba}$, any oriented orthonormal basis of
$\bSigma_a$, consisting of principal vectors.\vskip 0.5cm\par\noindent {\bf Theorem 3:}
{\it If $\{{\vt e}_\alpha,\alpha=1,2,3\}$, is a principal basis of $\phi_{ba}$, then the
vectors:} $${\vt e}_\alpha^\prime=:\lambda_{ba}({\vt e}_\alpha)^{-1}\sigma_{ba}{\vt
e}_\alpha\quad,\quad\alpha=1,2,3,\eqno(4.6)$$ {\it form an oriented, orthonormal basis in
$\bSigma_b$}.\par\noindent Proof. Since $\sigma_{ba}$ preserves the orientations, and
$\lambda_{ba}({\vt e}_\alpha)>0$, then $\{{\vt e}_\alpha^\prime,\alpha=1,2,3\}$, is an
oriented basis of $\bSigma_b$. Moreover, since the ${\vt e}_\alpha$'s are principal vectors,
then: $$\eqalign{h_b({\vt e}_\alpha^\prime,{\vt e}_\beta^\prime)&=\lambda_{ba}({\vt
e}_\alpha)^{-1}\lambda_{ba}({\vt e}_\beta)^{-1}h_b(\sigma_{ba}{\vt
e}_\alpha,\sigma_{ba}{\vt e}_\beta)=\cr &=\lambda_{ba}({\vt e}_\alpha)\lambda_{ba}({\vt
e}_\beta)^{-1}h_a({\vt e}_\alpha,{\vt e}_\beta)=\cr &=\delta_{\alpha\beta}.\cr}$$ Hence
$\{{\vt e}_\alpha^\prime,\alpha=1,2,3\}$ is orthonormal.\hfill\squarebox\par\hfill\vskip
0.5cm\par
Clearly, the same kinematical meanings are found for the coefficients $\Delta_{ab}$,
$\tau_{ab}$ and $\sigma_{ab}$ of the inverse transformation $\phi_{ab}$. Moreover, they are
correlated to $\Delta_{ba}$, $\tau_{ba}$ and $\sigma_{ba}$ by the cocycle relations (3.4).
These cocycle relations also lead to the following identities:
$$\eqalign{\sigma_{ba}(\hbox{\rm ker} \ \tau_{ba})&=\hbox{\rm ker} \ \tau_{ab}\cr \forall
\bxi\in\hbox{\rm ker} \ \tau_{ba} \ , \ \sigma_{ab}\circ\sigma_{ba}\bxi&=\bxi\cr
\forall\bxi\in\hbox{\rm ker} \ \tau_{ba} \ , \ \lambda_{ab}(\sigma_{ba}\bxi)&=
\lambda_{ba}(\bxi)^{-1}\cr \sigma_{ab}(\hbox{\rm ker} \ \tau_{ab})&=\hbox{\rm ker} \
\tau_{ba}\cr \forall\bxi^\prime\in\hbox{\rm ker} \ \tau_{ab} \ , \
\sigma_{ba}\circ\sigma_{ab}\bxi^\prime&=\bxi^\prime\cr \forall\bxi^\prime\in \hbox{\rm
ker} \ \tau_{ab} \ , \ \lambda_{ab}(\sigma_{ab}\bxi^\prime)&=
\lambda_{ab}(\bxi^\prime)^{-1}.\cr}\eqno(4.7)$$ Indeed, by $(3.4)_3$ and $(3.4)_7$ it
follows that, for any $\bxi\in\hbox{\rm ker} \ \tau_{ba}$, we have:
$$<\tau_{ab}\mid\sigma_{ba}\bxi>=-\Delta_{ab}<\tau_{ba}\mid\bxi>=0,$$ hence:
$\sigma_{ba}(\hbox{\rm ker} \ \tau_{ba})\subseteq\hbox{\rm ker} \ \tau_{ab}$; then, the
identity $(4.7)_1$ follows from: $\hbox{\rm dim ker} \ \tau_{ab}=\hbox{\rm dim ker} \
\tau_{ba}$. Moreover, by $(3.4)_4$ it follows that any $\bxi\in\hbox{\rm ker} \ \tau_{ba}$
also fulfils the equations: $$\eqalign{\bxi&=\sigma_{ab}\circ(\sigma_{ba}-\tau_{ba}
\otimes{\vt u}_{ba})\bxi=\cr
&=\sigma_{ab}\sigma_{ba}\bxi-<\tau_{ba}\mid\bxi>\sigma_{ab}{\vt u}_{ba}=\cr
&=\sigma_{ab}\sigma_{ba}\bxi,\cr}$$ and: $$\eqalign{\lambda_{ab}(\sigma_{ba}\bxi)&=
{h_a(\sigma_{ab}\sigma_{ba}\bxi,\sigma_{ab}\sigma_{ba}\bxi)^{1/2}
\over{h_b(\sigma_{ba}\bxi,\sigma_{ba}\bxi)^{1/2}}}=\cr &={h_a(\bxi,\bxi)^{1/2}
\over{h_b(\sigma_{ba}\bxi,\sigma_{ba}\bxi)^{1/2}}}=\cr
&={1\over{\lambda_{ba}(\bxi)}}.\cr}$$ The proof of identities $(4.7)_{4,5,6}$ is similar.\par
Let's now refer the space $\bSigma_a$ to a principal basis of $\phi_{ba}$, $\{{\vt e}_\alpha,
\alpha=1,2,3\}$, and refer the space $\bSigma_b$ to the oriented, orthonormal basis
defined by (4.6). By introducing the components of $\tau_{ba}$ and ${\vt
u}_{ab}$: $$\tau_\alpha=:<\tau_{ba}\mid{\vt e}_\alpha>\quad;\quad u^\alpha=:h_a({\vt
u}_{ab},{\vt e}_\alpha)\quad,\quad\alpha=1,2,3,$$ we derive, from (4.2), the following
coordinate expression of the space-time transformation $\phi_{ba}$:
$$\eqalign{t^\prime&=\Delta_{ba}t+\tau_1x^1+\tau_2x^2+\tau_3x^3\cr
x^{\prime\alpha}&=\lambda_{ba}({\vt e}_\alpha)(x^\alpha-u^\alpha t)
\qquad\alpha=1,2,3.\cr}\eqno(4.8)$$ Clearly, the inverse transformation formulas have not
the same simple structure of (4.8). This is due to the fact that, although the vectors ${\vt
e}^\prime_\alpha=\lambda_{ba}({\vt e}_\alpha)^{-1}\sigma_{ba}{\vt e}_\alpha$,
$\alpha=1,2,3$ form an oriented and orthonormal basis, they are not principal
vectors of the inverse transformation $\phi_{ab}$, in general.\vskip 1 cm\par\noindent
{\bf 5. The reciprocity conditions for a transformation.}\vskip 0.5cm\par
Formulas (4.8) are the simplest expression, in terms of coordinates, of the transformation
$\phi_{ba}$. In fact, no feature can be find in the cocycle relations (3.4) - i.e. in the
inertial equivalence relation - leading to a simpler expression. Such a
simplification only follows from some further assumptions correlating the space
transformation, $\sigma_{ba}$, and the simultaneity defect, $\tau_{ba}$, to the dragging
velocity, ${\vt u}_{ab}$, by means of the euclidean metrics of $\Sigma_a$ and $\Sigma_b$.
These assumptions, which we call the {\it reciprocity conditions}, are:\par\noindent
\item{1.} {\it The dragging velocity, ${\vt u}_{ab}$, is a principal vector of $\phi_{ba}$.}
\par\noindent
\item{2.} {\it The deformation ratio of space measures, $\lambda_{ba}$, is constant
on the orthogonal plane, $({\vt u}_{ab})^\perp$.}\par\noindent
\item{3.} {\it The simultaneity defect, $\tau_{ba}$, depends linearly on the dragging velocity:
$$\forall\bxi\in\bSigma_a\quad,\quad<\tau_{ba}\mid\bxi>=\theta h_a({\vt u}_{ab},\bxi),$$
where $\theta\in\reale$.}\par\noindent
Although not a consequence of the inertial equivalence relation, the reciprocity
conditions are taken in any theory of space and time. In fact, in a relativistic theory,
the reciprocity conditions are taken to describe all the space-time transformations, while
in a theory founded on the absolute frame, they are taken to describe only the space-time
transformations from the absolute frame to any other ``relative" frame.\par
The conditions 1 and 2 concern the structure of the spatial deformations affecting the
transformation $\phi_{ba}$. More precisely, the condition 1 says that the ratio:
$$\lambda=:\lambda_{ba}({\vt u}_{ab})=\Delta_{ba}{u_{ba}\over u_{ab}},$$ (where
$u_{ba}=:\mid{\vt u}_{ba}\mid$, $u_{ab}=:\mid{\vt u}_{ab}\mid$), is a principal deformation
ratio of $\phi_{ba}$. Condition 1 have the following kinematical meaning: any rod at rest
in $b$, which is orthogonal to the dragging velocity ${\vt u}_{ba}$, have instantaneous
configurations, in $a$, which are orthogonal to the dragging velocity ${\vt u}_{ab}$.\vskip
0.5cm\par\noindent
{\bf Theorem 4:} {\it The condition 1 is equivalent to:} $$\sigma_{ba}({\vt
u}_{ab})^\perp=({\vt u}_{ba})^\perp.\eqno(5.1)$$
Proof. If ${\vt u}_{ab}$ is assumed to be principal, then any $\bxi\in({\vt u}_{ab})^\perp$
fulfils: $$\eqalign{h_b(\sigma_{ba}\bxi,{\vt u}_{ba})&=
-\Delta_{ba}^{-1}h_b(\sigma_{ba}\bxi,\sigma_{ba}{\vt u}_{ab})=\cr
&=-\Delta_{ba}^{-1}\lambda^2h_a(\bxi,{\vt u}_{ab})=\cr &=0,\cr}$$ hence: $\sigma_{ba}({\vt
u}_{ab})^\perp\subseteq({\vt u}_{ba})^\perp$. Therefore, since dim $\sigma_{ba}({\vt
u}_{ab})^\perp=2=$dim $({\vt u}_{ba})^\perp$, then we find (5.1). Conversely, if the equation
(5.1) is assumed to hold, then by decomposing any $\bxi\in\bSigma_a$ as: $$\bxi=c{\vt
u}_{ab}+\bxi^\perp\quad,\quad\bxi^\perp\in({\vt u}_{ab})^\perp,$$ we find that:
$$\eqalign{h_b(\sigma_{ba}\bxi,\sigma_{ba}{\vt u}_{ab})&=ch_b(\sigma_{ba}{\vt u}_{ab},
\sigma_{ba}{\vt u}_{ab})+h_b(\sigma_{ba}\bxi^\perp,\sigma_{ba}{\vt u}_{ab})=\cr
&=c\lambda^2h_a({\vt u}_{ab},{\vt u}_{ab})-\Delta_{ba}h_b(\sigma_{ba}\bxi^\perp,{\vt
u}_{ba})=\cr &=\lambda^2h_a(c{\vt u}_{ab},{\vt u}_{ab})=\cr &=\lambda^2h_a(c{\vt
u}_{ab}+\bxi^\perp,{\vt u}_{ab})=\cr &=\lambda^2h_a(\bxi,{\vt u}_{ab}),\cr}$$ i.e. ${\vt
u}_{ab}$ is a principal vector of $\phi_{ba}$.\hfill\squarebox\par\hfill\vskip 0.5cm\par
Condition 2 means that the rods at rest in $b$ whose instantaneous
configurations in $a$ are orthogonal to the dragging velocity ${\vt u}_{ab}$, all define the
same deformation ratio of space measures. Therefore, no direction in the plane
$({\vt u}_{ab})^\perp$ is {\it privileged} by the space-time transformation. 
Condition 2, together with condition 1, implies that this constant ratio:
$$\mu=:\lambda_{ba}(\bxi)\quad,\quad\bxi\in({\vt u}_{ab})^\perp,$$ is a principal
deformation ratio and, then, that the orthogonal plane, $({\vt u}_{ab})^\perp$, is a {\it
principal subspace} of $\phi_{ba}$. A further kinematical meaning of condition 2 is
founded in the following theorem.\vskip 0.5cm\par\noindent 
{\bf Theorem 5:} {\it If ${\vt u}_{ab}$ is a principal vector, then condition 2 is valid if and
only if the restricted space transformation: $$\sigma_{ba}:({\vt
u}_{ab})^\perp\to({\vt u}_{ba})^\perp$$ is a conformal mapping.}\par\noindent
Proof. Clearly, the restricted mapping is conformal if and only if the following identity holds:
$$h_b(\sigma_{ba}\bxi,\sigma_{ba}\bfeta)=\lambda_{ba}(\bxi)\lambda_{ba}(\bfeta)h_a
(\bxi,\bfeta),\eqno(5.2)$$ for any couple of nonzero vectors, $\bxi,\bfeta\in({\vt
u}_{ab})^\perp$. Now, if condition 2 is true, then $({\vt u}_{ab})^\perp$ is a principal
subspace, and $\lambda_{ba}$ is constant on it; hence:
$$\eqalign{h_b(\sigma_{ba}\bxi,\sigma_{ba}\bfeta)&=\mu^2h_a(\bxi,\bfeta)=\cr
&=\lambda_{ba}(\bxi)\lambda_{ba}(\bfeta)h_a(\bxi,\bfeta).\cr}$$ Conversely, let's assume
that (5.2) holds - together with the condition 1. Now, if $\bfeta\in({\vt u}_{ab})^\perp$ is
any principal vector of $\phi_{ba}$, then any nonzero vector $\bxi\in({\vt u}_{ab})^\perp$
obeys the relation: $$h_b(\sigma_{ba}\bxi,\sigma_{ba}\bfeta)=\lambda_{ba}(\bfeta)^2h_a
(\bxi,\bfeta).$$ Moreover, if $\bxi$ isn't orthogonal to $\bfeta$, then by (5.2) it follows that
$\sigma_{ba}\bxi\in({\vt u}_{ba})^\perp$ isn't orthogonal to $\sigma_{ba}\bfeta$, and that:
$$\lambda_{ba}(\bxi)=\lambda_{ba}(\bfeta).$$ From this we infer that $\lambda_{ba}$ is
constant on $({\vt u}_{ab})^\perp$.\hfill\squarebox\par\hfill\vskip 0.5cm\par
Conditions 1 and 2 imply that any oriented orthonormal basis of $\bSigma_a$ of the type:
$${\vt e}_1=:u^{-1}_{ab}{\vt u}_{ab}\quad;\quad{\vt e}_\alpha\in({\vt
u}_{ab})^\perp\quad,\quad\alpha=2,3,\eqno(5.3)$$ is a principal basis of $\phi_{ba}$.
Moreover, the oriented orthonormal basis of $\bSigma_b$ defined in (4.6) is given by:
$${\vt e}^\prime_1=\lambda^{-1}\sigma_{ba}{\vt e}_1\quad;\quad{\vt
e}^\prime_\alpha=\mu^{-1}\sigma_{ba}{\vt e}_\alpha\quad,\quad\alpha=2,3,$$
and fulfils: $${\vt e}^\prime_1=-u^{-1}_{ba}{\vt u}_{ba}\quad;\quad{\vt
e}^\prime_\alpha\in({\vt u}_{ba})^\perp\quad,\quad\alpha=2,3.\eqno(5.4)$$
Now, referred to the bases (5.3) and (5.4), the coordinate expression of the space-time
transformation $\phi_{ba}$ reduces to the simpler form: $$\eqalign{t^\prime&=\Delta_{ba}t+
\tau_1 x^1+\tau_2 x^2+\tau_3 x^3\cr {x^\prime}^1&=\lambda(x^1-u_{ab}t)\cr
{x^\prime}^\alpha&=\mu x^\alpha\quad,\quad\alpha=2,3.\cr}$$
The condition 3 concerns the simultaneity defect affecting the space-time transformation
$\phi_{ba}$. It means that simultaneity be conserved, in the transformation
$\phi_{ba}$, for any couple of events which are spatially separated, in $a$, by a vector which
is orthogonal to the dragging velocity, ${\vt u}_{ab}$. Indeed, if the events $x_1,x_2\in M$
are simultaneous with respect to $a$, then their time separation, measured in $b$, is:
$$t^\prime_2-t^\prime_1=<\tau_{ba}\mid{\vt r}_2-{\vt r}_1>=\theta h_a({\vt u}_{ab},{\vt
r}_2-{\vt r}_1).$$ Hence, if ${\vt r}_2-{\vt r}_1\in({\vt u}_{ab})^\perp$, then:
$t^\prime_2=t^\prime_1$.\par
The condition 3 is manifestly equivalent to the inclusion: $$({\vt u}_{ab})^\perp\subseteq
\hbox{\rm ker} \ \tau_{ba}.\eqno(5.5)$$ Clearly, inclusion (5.5) becames an identity, if
$\tau_{ba}\not=0$. Moreover, in any basis of the type (5.3), the components of $\tau_{ba}$
are: $$\tau_1=u_{ab}\theta\quad;\quad\tau_2=\tau_3=0.$$\par
We now suppose that all the reciprocity conditions hold for the space-time
transformation $\phi_{ba}$. Then the principal bases and the subspace of conserved
simultaneity of $\phi_{ba}$ depend only on the dragging velocity, ${\vt u}_{ab}$, by means
of (5.3) and (5.5). Moreover, the coordinate expression of $\phi_{ba}$, referred to the bases
(5.3) and (5.4), reduces to the simplified form:
$$\eqalign{t^\prime&=\Delta_{ba}t+u_{ab}\theta x^1\cr
{x^\prime}^1&=\lambda(x^1-u_{ab}t)\cr {x^\prime}^\alpha&=\mu x^\alpha\quad,
\quad\alpha=2,3.\cr}\eqno(5.6)$$ Formulas (5.6) are the simplest
expression in terms of coordinates of the space-time transformation $\phi_{ba}$, as a
consequence of the reciprocity conditions. We call them the {\it special coordinate
expression} of $\phi_{ba}$. Formulas (5.6) shows that the transformation $\phi_{ba}$ is
completely determined by the dragging velocity, ${\vt u}_{ab}$, and the four parameters:
$$\Delta_{ba},\lambda,\mu\in\reale^+\quad;\quad\theta\in\reale.\eqno(5.7)$$ Moreover,
it is a simple matter to show that, if a space-time transformation $\phi_{ba}$ has a special
coordinate expression, then its coefficients must satisfy the reciprocity conditions.\par
An important feature of the reciprocity conditions is that they are symmetric under the
exchange of the two reference frames involved in the space-time transformation.\vskip
0.5cm\par\noindent
{\bf Theorem 6:} {\it The reciprocity conditions 1-3 are fulfilled by the coefficients of the
inverse transformation, $\phi_{ab}$, if and only if they are fulfilled by the coefficients of
$\phi_{ba}$.}\par\noindent
Proof. It is sufficient to restrict our proof to the ``if" part of the theorem. Let's assume that
$\phi_{ba}$ fulfils the reciprocity conditions. To see that the inverse transformation,
$\phi_{ab}$, also fulfils these conditions, we must show that the coefficients of
$\phi_{ab}$ fulfil the equations: $$\sigma_{ab}({\vt u}_{ba})^\perp=({\vt
u}_{ab})^\perp\eqno(5.1)^\prime$$ $$\forall\bxi^\prime,\bfeta^\prime\in({\vt
u}_{ba})^\perp\quad,\quad
h_a(\sigma_{ab}\bxi^\prime,\sigma_{ab}\bfeta^\prime)=\lambda_{ab}(\bxi^\prime)
\lambda_{ab}(\bfeta^\prime)h_b(\bxi^\prime,\bfeta^\prime)\eqno(5.2)^\prime$$ $$({\vt
u}_{ba})^\perp\subseteq\hbox{\rm ker} \ \tau_{ab},\eqno(5.5)^\prime$$ Indeed, from (5.5)
and $(4.7)_2$ we find that: $$\sigma_{ab}\circ\sigma_{ba}\mid_{({\vt u}_{ab})^\perp}=
\hbox{\rm id};$$ hence, by (5.1) it follows that: $$\sigma_{ab}({\vt u}_{ba})^\perp=
\sigma_{ab}\circ\sigma_{ba}({\vt u}_{ab})^\perp=({\vt u}_{ab})^\perp.$$ Moreover, by (5.1)
and $(4.7)_2$, it follows that any pair of nonzero vectors, $\bxi^\prime,\bfeta^\prime\in
({\vt u}_{ba})^\perp$, fulfil:
$$\eqalign{h_a(\sigma_{ab}\bxi^\prime,\sigma_{ab}\bfeta^\prime)&=
h_a(\sigma_{ab}\sigma_{ba}\bxi,\sigma_{ab}\sigma_{ba}\bfeta)=\cr
&=h_a(\bxi,\bfeta),\cr}$$ where $\bxi^\prime=\sigma_{ba}\bxi$,
$\bfeta^\prime=\sigma_{ba}\bfeta$. Now, by $(4.7)_3$ and (5.2) we find:
$$\eqalign{h_a(\sigma_{ab}\bxi^\prime,\sigma_{ab}\bfeta^\prime)&=h_a(\bxi,\bfeta)=\cr
&=\lambda_{ba}(\bxi)^{-1}\lambda_{ba}(\bfeta)^{-1}h_b(\sigma_{ba}\bxi,\sigma_{ba}\bfeta)
=\cr &=\lambda_{ab}(\sigma_{ba}\bxi)\lambda_{ab}(\sigma_{ba}\bfeta)h_b(\sigma_{ba}\bxi,
\sigma_{ba}\bfeta)=\cr
&=\lambda_{ab}(\bxi^\prime)\lambda_{ab}(\bfeta^\prime)h_b(\bxi^\prime,\bfeta^\prime).
\cr}$$ Finally, by (5.1), (5.5) and $(4.7)_1$ we find: $$({\vt u}_{ba})^\perp=\sigma_{ba}({\vt
u}_{ab})^\perp\subseteq\sigma_{ba}(\hbox{\rm ker} \ \tau_{ba})=\hbox{\rm ker} \
\tau_{ab}.$$\hfill\squarebox\par\hfill\vskip 0.5cm\par
Clearly, the inverse of any special coordinate expression of the transformation $\phi_{ba}$ is
a special coordinate expression of the inverse transformation, $\phi_{ab}$. Indeed, the
principal deformation ratios affecting $\phi_{ab}$ are: $$\lambda^\prime=\lambda_{ab}({\vt
u}_{ba})=\Delta_{ab}{u_{ab}\over u_{ba}}\quad;\quad\mu^\prime={1\over\mu};$$ in fact, for
any $\bxi^\prime\in({\vt u}_{ba})^\perp$, we find: $\sigma_{ab}\bxi^\prime\in({\vt
u}_{ab})^\perp$, and then: $$\mu^\prime=\lambda_{ab}(\bxi^\prime)=
\lambda_{ab}(\sigma_{ba}\sigma_{ab}\bxi^\prime)={1\over{\lambda_{ba}
(\sigma_{ab}\bxi^\prime)}}={1\over\mu}.$$ Moreover, the basis (5.4) is a principal basis of
$\phi_{ab}$, and its associated basis in $\bSigma_a$ by means of (4.6) is nothing but the
basis (5.3): $${\lambda^\prime}^{-1}\sigma_{ab}{\vt e}^\prime_1={\vt
e}_1\quad;\quad{\mu^\prime}^{-1}\sigma_{ab}{\vt e}^\prime_\alpha={\vt
e}_\alpha\quad,\quad\alpha=2,3.$$ Therefore, the coordinate expression of $\phi_{ab}$,
referred to the bases (5.4) and (5.3), consists of the inverse of (5.6). Since:
$$\tau^\prime_1=<\tau_{ab}\mid{\vt
e}^\prime_1>=-u_{ba}\theta^\prime\quad;\quad\tau^\prime_\alpha=<\tau_{ab}\mid{\vt
e}^\prime_\alpha>=0\quad,\quad\alpha=2,3,$$ then this coordinate expression is:
$$\eqalign{t&=\Delta_{ab}t^\prime-u_{ba}\theta^\prime{x^\prime}^1\cr
x^1&=\lambda^\prime({x^\prime}^1+u_{ba}t^\prime)\cr
x^\alpha&=\mu^\prime{x^\prime}^\alpha\quad,\quad\alpha=2,3.\cr}\eqno(5.8)$$
Finally, the dragging velocity $u_{ba}$ and the parameters $\Delta_{ab},\lambda^\prime,
\mu^\prime,\theta^\prime$, can be expressed as functions of $u_{ab},\Delta_{ba},\lambda,
\mu$ and $\theta$, by means of:
$$\eqalign{u_{ba}&={\lambda\over\Delta_{ba}}u_{ab};\cr
\Delta_{ab}&={1\over{\Delta_{ba}+u_{ab}^2\theta}}\quad;\quad\lambda^\prime={\Delta_{ba}
\over{\lambda(\Delta_{ba}+u_{ab}^2\theta)}}\quad;\quad\mu^\prime={1\over\mu}\quad;\quad
\theta^\prime={\Delta_{ba}\theta\over{\lambda^2(\Delta_{ba}+u_{ab}^2\theta)}}.\cr}$$
\vfill\eject\par\noindent
{\bf 6. Theories of space and time.}\vskip 0.5cm\par
As we have remarked in section 3, the inertia principle, mathematically, consists of the
singling out of a class of inertially equivalent reference frames, $I$. This class contains all
the reference frames in which the free material particles are uniformly
moving.\par
The aim of the present section is to suggest a definition of what can be meant
by a theory of space and time compatible with the inertia principle. To this end, beside the
class $I$, a further mathematical concept must be introduced to determine the space-time
transformations between the inertial frames of reference. This object, which we call the
{\it rest frame map family}, consists of a family of mappings: $${\cal R}=:\{{\cal R}_a:C_a\to
I/a\in I\},$$ with the following properties:\par\noindent
1. {\it Any ${\cal R}_a:C_a\to I$ is one-to-one};\par\noindent
2. $\forall a,b\in I,\quad{\cal R}_a^{-1}(b)={\vt u}_{ab}$;\par\noindent
3. $\forall a,b\in I,\quad{\cal R}_b^{-1}\circ{\cal R}_a=\Phi_{ba}$.\par
Properties 1 and 2 mean that if an inertial frame, $a\in I$, is fixed, then any other frame of
the class $I$ can be found, in $a$, as the rest frame associated to some free material
particle. Moreover, since the mapping ${\cal R}_a$ uniquely determines any $b\in I$ as a
function of the dragging velocity ${\vt u}_{ab}$, then the space-time transformation
$\phi_{ba}$ is determined as a function of the dragging velocity. Therefore, the
coefficients of $\phi_{ba}$ are found, by means of ${\cal R}_a$, as ``constitutive" functions
of the dragging velocity. To state this more precisely, let's denote by
$L_{ab}\subset\bSigma_a^*\otimes \bSigma_b$ the set of orientation preserving linear
isomorphisms from $\bSigma_a$ to $\bSigma_b$ and costruct, for a fixed $a\in I$, the
bundle of orientation preserving linear isomorphisms over $I$ by means of:
$$L_a=:\cup_{b\in I}L_{ab}$$ with the bundle projection, $\pi_a:L_a\to I,$ defined by:
$\pi_a(\sigma)=b\Leftrightarrow:\sigma\in L_{ab}$. Now, the rest frame mapping ${\cal
R}_a:C_a\to I$ uniquely determines the {\it constitutive functions}:
$$\Delta^{(a)}:C_a\to\reale^+\quad;\quad\tau^{(a)}:C_a\to\bSigma_a^*\quad;\quad
\sigma^{(a)}:C_a\to L_a,\eqno(6.1)$$ by means of the equation:
$$d\phi_{ba}=:\pmatrix{\Delta^{(a)}({\vt u}){\vt e}_0^*\otimes{\vt
e}_0^\prime&\tau^{(a)}({\vt u})\otimes{\vt e}_0^\prime\cr -{\vt
e}^*_0\otimes\sigma^{(a)}({\vt u}){\vt u}&\sigma^{(a)}({\vt u})\cr},$$ where ${\vt u}={\vt
u}_{ab}$. The functions (6.1) must satisfy the properties:
$$\pi_a\circ\sigma^{(a)}={\cal R}_a,\eqno(6.2)$$ $$\Delta^{(a)}({\vt
0})=1\quad;\quad\tau^{(a)}({\vt 0})=0\quad;\quad\sigma^{(a)}({\vt 0})=\hbox{\rm
id}_{\bSigma_a}.\eqno(6.3)$$\par
Property 3 is nothing but a compatibility condition between the mappings of ${\cal R}$. It
implies that if a rest frame mapping, ${\cal R}_a:C_a\to I$, is given, then any other mapping
of the family ${\cal R}$ can be deduced from ${\cal R}_a$ via the cocycle relations (3.4) and
(3.5). More precisely, if $b\in I$ is any inertial frame, then the constitutive functions
determined by the rest frame mapping ${\cal R}_b:C_b\to I$:
$$\Delta^{(b)}:C_b\to\reale^+\quad;\quad\tau^{(b)}:C_b\to\bSigma_b^*\quad;\quad
\sigma^{(b)}:C_b\to L_b,$$ must be related to the functions (6.1) by the equations:
$$\eqalign{\Delta^{(b)}({\vt v}^\prime)&=(\Delta^{(a)}({\vt u})+<\tau^{(a)}({\vt u})
\mid{\vt u}>)^{-1}(\Delta^{(a)}({\vt v})+<\tau^{(a)}{\vt v})\mid{\vt u}>)\cr
\tau^{(b)}({\vt v}^\prime)&=(\Delta^{(a)}({\vt u})\tau^{(a)}({\vt v})-\Delta^{(a)}({\vt v})
\tau^{(a)}({\vt u}))\circ\cr
&\quad\quad\quad\quad\quad\quad\quad\quad\quad\quad\circ(\Delta^{(a)}({\vt
u})\sigma^{(a)}({\vt u})+\tau^{(a)}({\vt u})\otimes\sigma^{(a)}({\vt u}){\vt u})^{-1}\cr
\sigma^{(b)}({\vt v}^\prime)&=\sigma^{(a)}({\vt v})\circ(\Delta^{(a)}({\vt
u})\hbox{\rm id}_{\bSigma_a}+\tau^{(a)}({\vt u})\otimes({\vt v}))\circ\cr
&\quad\quad\quad\quad\quad\quad\quad\quad\quad\quad\circ
(\Delta^{(a)}({\vt u})\sigma^{(a)}({\vt u})+\tau^{(a)}({\vt u})\otimes
\sigma^{(a)}({\vt u}){\vt u})^{-1},\cr}\eqno(6.4)$$ where ${\vt v}^\prime\in C_b$, ${\vt
u}={\vt u}_{ab}={\cal R}_a^{-1}(b)$ and ${\vt v}=\Phi_{ab}({\vt v}^\prime)$.\par
The assignment of a rest frame map family on the class $I$ is completely equivalent to the
assignment of the constitutive functions (6.1) for any $a\in I$. Indeed, it is a simple matter
to prove the following theorem:\vskip 0.5cm\par\noindent
{\bf Theorem 7}: {\it Let $I$ be a class of inertially equivalent reference frames, and let:
$$\{\Delta^{(a)},\tau^{(a)},\sigma^{(a)}/a\in I\}$$ be a family of functions of the type (6.1). If
these functions fulfil the properties (6.3) and (6.4) and if, in addition, the mappings: $${\cal
R}_a=:\pi_a\circ\sigma^{(a)}:C_a\to I\quad,\quad a\in I$$ are one-to-one, then ${\cal R}$
is a rest frame map family based on $I$.}\hfill\squarebox\par\hfill\vskip 0.5cm\par
We conclude by defining a {\it theory of space and time compatible with the inertia
principle} as a triplet, $${\cal T}=:(M;I;{\cal R}),$$ consisting of: the universe of events,
$M$; a class of inertially equivalent reference frames of $M$, $I$; a rest frame
map family based on $I$, ${\cal R}$. All the theories of space and time which are compatible
with the inertia principle can be obtained, as examples of this definition, by specifying the
rest frame map family ${\cal R}$ - i.e. by specifying the constitutive functions (6.1).\vskip
0.5cm\par\noindent
{\bf Example 1:} {\bf Galileian relativity and special relativity}. Let's suppose that a theory
${\cal T}$ is such that, for a given frame $a\in I$, all the space-time transformations
$\phi_{ba}$ fulfil the reciprocity conditions of section 5. Now, since any $\phi_{ba}$ is
determined by the four parameters in (5.7), then the constitutive functions of the rest frame
mapping ${\cal R}_a$ are determined by specifying the four real-valued functions:
$$\Delta^{(a)},\lambda^{(a)},\mu^{(a)}:C_a\to\reale^+\quad;\quad\theta^{(a)}:C_a\to\reale,
\eqno(6.5)$$ with the property that: $$\Delta^{(a)}({\vt 0})=\lambda^{(a)}({\vt
0})=\mu^{(a)}({\vt 0})=:1\quad;\quad \theta^{(a)}({\vt 0})=:0.$$ Now, Galilean relativity is
obtained by assuming: $$C_a=:\bSigma_a\eqno(6.6)$$ and $$\eqalign{\Delta^{(a)}({\vt
u})=\lambda^{(a)}({\vt u})=\mu^{(a)}({\vt u})&=:1\cr \theta^{(a)}({\vt
u})&=:0.\cr}\eqno(6.7)$$ It is a simple matter to show, as a consequence of (6.4), that
reciprocity conditions and equations (6.6)-(6.7) hold for any other reference frame of the
class $I$. Similarly, special relativity is obtained by assuming:
$$C_a=:\{{\vt u}\in\bSigma_a/u<c\}\eqno(6.8)$$ and: $$\eqalign{\Delta^{(a)}({\vt
u})=\lambda^{(a)}({\vt u})&=:\left(1-{u^2\over c^2}\right)^{-1/2}\cr
\mu^{(a)}({\vt u})&=:1\cr \theta^{(a)}({\vt u})&=:-{1\over c^2}\left(1-{u^2\over
c^2}\right)^{-1/2}.\cr}\eqno(6.9)$$ Also for these functions it is posible to show, as a
consequence of (6.4), that reciprocity conditions and equations (6.6)-(6.7) hold for any
other reference frame of the class $I$.\vskip 0.5cm\par\noindent
{\bf Example 2:} {\bf Lorentz relativity and the absolute theory kinematically equivalent to
special relativity}. Any theory ${\cal T}$, based on the existence of the absolute frame,
$a\in I$, can be obtained by assuming, as in the previous example, that all the
transformations $\phi_{ba}$ fulfil the reciprocity. Now the Lorentz's relativity [23]
is obtained by assuming: $$C_a=:\{{\vt u}\in\bSigma_a/u<c\}\eqno(6.10)$$ and
$$\eqalign{\Delta^{(a)}({\vt u})=\lambda^{(a)}({\vt u})&=:\gamma(u)\left(1-{u^2\over
c^2}\right)^{-1/2}\cr \mu^{(a)}({\vt u})&=:\gamma(u)\cr \theta^{(a)}({\vt
u})&=:-{1\over c^2}\gamma(u)\left(1-{u^2\over c^2}\right)^{-1/2},\cr}\eqno(6.11)$$ where
$\gamma:[0,c)\to\reale^+$ is a monotonic function such that $\gamma(0)=1$; while the
absolute theory kinematically equivalent to special relativity [22], [25-27], [29], [30], is
obtained by assuming $C_a$ as in (6.10) and: $$\eqalign{\Delta^{(a)}({\vt
u})&=:\left(1-{u^2\over c^2} \right)^{1/2}\cr \lambda^{(a)}({\vt u})&=:\left(1-{u^2\over
c^2}\right)^{-1/2}\cr \mu^{(a)}({\vt u})&=:1\cr \theta^{(a)}({\vt
u})&=:0.\cr}\eqno(6.12)$$ It is a consequence of equations (6.4) that in both these theories
the reciprocity conditions for a space-time transformation $\phi_{cb}$, with $c\not=a,
b\not=a$, and such that ${\vt u}_{ac}$ is not proportional to ${\vt u}_{ab}$, are not fulfilled.
\vskip 1cm\par\noindent
{\bf Aknowledgements.}\vskip 0.5 cm\par\noindent
The original idea of this paper has been suggested us by many stimulating discussions with
Franco Selleri. His encouraging support is aknowledged. Further, we would like to thank A.
Afriat, U. Bartocci and M. Mamone Capria for helping us to improve the paper with their
fruitful comments.\vskip 1cm\par\noindent
{\bf References.}\vskip 0.5 cm\par\noindent
\item{[1]} J. B. Barbour, ABSOLUTE OR RELATIVE MOTION?, Cambridge Univ. Press, Cambridge,
1989.
\item{[2]} P. W. Bridgman, THE LOGIC OF MODERN PHYSICS, The Macmillan Company, New
York, 1927.
\item{[3]} E. Cassirer, ZUR EINSTEIN'SCHEN RELATIVITATSTHEORIE, Bruno
Cassirer Verlag, Berlin, 1920.
\item{[4]} I. Ciufolini, J. A. Wheeler, GRAVITATION AND INERTIA, Princeton Univ. Press,
Princeton, 1995.
\item{[5]} A. Gr\"unbaum, PHILOSOPHICAL PROBLEMS OF SPACE AND TIME, Reidel,
Dor\-dre\-cht,
1973.
\item{[6]} M. Jammer, CONCEPTS OF SPACE. THE HISTORY OF THE THEORIES OF SPACE IN
PHYSICS, Harvard Univ. Press, Cambridge, Mass., 1954.
\item{[7]} A. Koir\'e, NEWTONIAN STUDIES, Harvard Univ. Press, Cambridge, Mass., 1965.
\item{[8]} A. Koir\'e, ETUDES GALILEENNES, Hermann, Paris, 1966.
\item{[9]} P. J. E. Peebles, PRINCIPLES OF PHYSICAL COSMOLOGY, Princeton Univ. Press,
Princeton, 1993.
\item{[10]} H. Reichenbach, THE PHILOSOPHY OF SPACE AND TIME, Dover, New York, 1958.
\item{[11]} M. Schlick, RAUM UND ZEIT IN DER GEGENWARTIGEN PHYSIK, Springer Verlag,
Berlino,1922.
\item{[12]} R. Torretti, PHILOSOPHY OF GEOMETRY FROM RIEMANN TO POINCARE', Reidel,
Dordrecht, 1978.
\item{[13]} R. Torretti, RELATIVITY AND GEOMETRY, Pergamon Press, Oxford, 1983.
\item{[14]} A. Trautman, in A. Trautman, F. A. E. Pirani, H. Bondi, LECTURES ON
GENERAL RELATIVITY, Brandeis 1964 Summer Institute on Theoretical Physics,
Prentice-Hall, Englewood Cliff, N. J., 1965.
\item{[15]} H. Weyl, SPACE-TIME-MATTER, Dover, New York, 1952.
\item{[16]} H. Weyl, PHILOSOPHY OF MATHEMATICS AND NATURAL SCIENCE, Dover, New York,
1963.
\item{[17]} E. Whittaker, A HISTORY OF THE THEORIES OF AETHER AND ELECTRICITY,
Dover, New York, 1989.
\item{[18]} V. Berzi, V. Gorini, {\it Reciprocity principle and Lorentz transformations}, J.
Math. Phys., {\bf 10} (1969),1518-1524.
\item{[19]} A. Einstein, {\it Zur Elektrodynamik bewegter K\"orper}, Ann. Phys., {\bf 17}
(1905), 891-921.
\item{[20]} P. Havas, {\it Four dimensional formulations of Newtonian
mechanics and their relation to the special and the general theory of relativity}, Rev. Mod.
Phys., {\bf 36} (1964), 938-965.
\item{[21]} H. P. K\"unzle, {\it Galilei and Lorentz structures in space-time: Comparison of
the corresponding geometry and physics}, Ann. Inst. H. Poincar\'e, {\bf XVII} (1972), 337-362.
\item{[22]} R. Mansouri, R. U. Sexl, {\it A test theory of special relativity}, I,II,III, Gen. Rel.
Grav., {\bf 8}, (1977), 497-537.
\item{[23]} J. H. Poincar\'e, {\it Sur la dynamique de l'electron}, Rend. Circ. Mat. Palermo, {\bf
21}, (1906), 129-175.
\item{[24]} W. A. Rodrigues Jr., Q. A. E. de Souza, Y. Bozhkov, Found. Phys., {\bf 25}, (1995),
871-924.
\item{[25]} F. Selleri, {\it Space, time and their transformations}, Chin. J. Eng. Electron., {\bf
6}, (1995), 24-44.
\item{[26]} F. Selleri, {\it Noninvariant one-way velocity of light}, Found. Phys., {\bf 26},
(1996), 641-664.
\item{[27]} F. Selleri, {\it Noninvariant one-way velocity of light and particle collision}, 
Found. Phys. Lett., {\bf 9}, (1996), 43-60.
\item{[28]} G. Spavieri, {\it Nonequivalence of ether theories and special relativity}, Phys.
Rev. A, {\bf 34}, (1986), 1708-1713.
\item{[29]} A. A. Ungar, {\it Formalism to deal with Reichenbach's special theory of
relativity}, Found. Phys., {\bf 21}, (1991), 691-726.
\item{[30]} Y. Z. Zhang, {\it Test theories of special relativity}, Gen. Rel. Grav., {\bf 27},
(1995), 475-493.\end